\documentclass[runningheads]{llncs}
\usepackage[T1]{fontenc}
\usepackage{graphicx}
\usepackage{caption}
\usepackage{subcaption}
\usepackage{natbib}
\usepackage{booktabs}
\usepackage{multirow}
\setcitestyle{square}
\let\svthefootnote\thefootnote

\begin{document}
\pagenumbering{gobble} 
\title{Opportunities of Reinforcement Learning in South Africa’s Just Transition}
%
%

\author{
Claude Formanek\inst{1}\and
Callum Rhys Tilbury\and
Jonathan P. Shock\inst{1,2,3}
}
\authorrunning{Formanek et al.}
%

\institute{Shocklab, University of Cape Town, South Africa
\and
INRS, Montreal, Canada
\and
NiTheCS, Stellenbosch, South Africa
}

\maketitle 

\let\thefootnote\relax\footnote{Preprint. Accepted at SACAIR 2024.}
\addtocounter{footnote}{-1}\let\thefootnote\svthefootnote

\begin{abstract}
South Africa stands at a crucial juncture, grappling with interwoven socio-economic challenges such as poverty, inequality, unemployment, and the looming climate crisis. The government's Just Transition framework aims to enhance climate resilience, achieve net-zero greenhouse gas emissions by 2050, and promote social inclusion and poverty eradication. According to the Presidential Commission on the Fourth Industrial Revolution, artificial intelligence technologies offer significant promise in addressing these challenges. This paper explores the overlooked potential of Reinforcement Learning (RL) in supporting South Africa's Just Transition. It examines how RL can enhance agriculture and land-use practices, manage complex, decentralised energy networks, and optimise transportation and logistics, thereby playing a critical role in achieving a just and equitable transition to a low-carbon future for all South Africans. We provide a roadmap as to how other researchers in the field may be able to contribute to these pressing problems.


\keywords{Reinforcement Learning \and Just Transition \and 4th Industrial Revolution \and South Africa \and Precision Agriculture \and Smart Grids \and Climate Change \and Smart Cities \and Artificial Intelligence }
\end{abstract}

\section{Introduction}
South Africa is at a pivotal moment in its history, facing a myriad of interlocking socio-economic challenges, primarily poverty, inequality, joblessness, and a looming climate crisis that threatens to exacerbate these issues. In response to these pressing concerns, the South African government has developed the Just Transition (JT) framework~\cite{justtransition}. This framework aims to achieve a high quality of life for all South Africans by increasing the country's ability to adapt to the adverse impacts of climate change, fostering climate resilience, and reaching net-zero greenhouse gas emissions by 2050. Simultaneously, it emphasises decent work for all, social inclusion, and eradicating poverty, with a particular focus on empowering the poor, women, people with disabilities, and the youth.

Central to the JT is the development of innovative technologies that enhance efficiency and resilience within the economy and society. The South African government's Commission on the Fourth Industrial Revolution (4IR)  \cite{gov20204ir} has explored the potential of new and emerging technologies, underscoring Artificial Intelligence (AI) as a disruptive force capable of transforming industries and our society. As a core pillar of the 4IR, AI stands to play a significant role in realising South Africa's JT.

\begin{figure}[t]
    \centering
    \begin{subfigure}{0.40\textwidth}
        \centering
        \includegraphics[width=\linewidth]{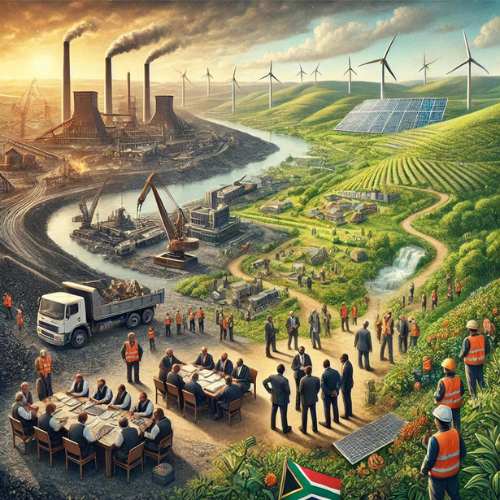}
        \caption{Just Transition}
    \end{subfigure}
    \hfill
    \begin{subfigure}{0.40\textwidth}
        \centering
        \includegraphics[width=\linewidth]{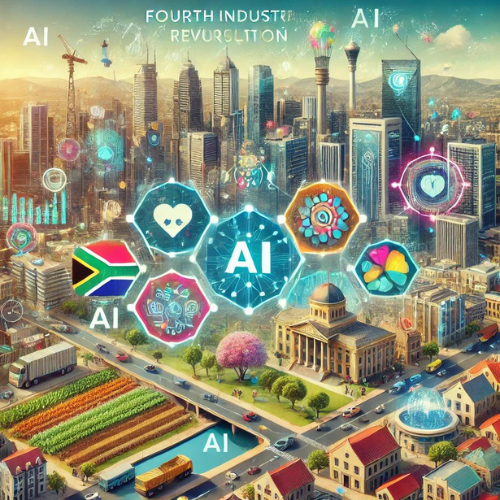}
        \caption{Fourth Industrial Revolution}
    \end{subfigure}
    \caption{On the left is a visual depiction of the Just Transition created by Generative AI model DALL-E~\cite{ramesh2021zeroshottexttoimagegeneration}. On the right, a depiction of the Fourth Industrial Revolution in South Africa was also generated using DALL-E~. We prompted DALL-E using the information in the background section of this report.}
\end{figure}


This paper explores how Reinforcement Learning (RL) \cite{sutton2018reinforcement}, one form of AI, can play a significant role in enhancing climate resilience and promoting sustainable and inclusive development. 
We conduct a systematic literature review of how RL has been leveraged globally to enhance our agricultural land-use practices to ensure food security and sustainable growth; manage decentralised and resilient renewable energy networks; and optimise complex and fragile transportation and logistic networks. Finally, we provide a roadmap for future research that extends the existing literature to the South African context. In doing so, this paper underscores the role RL could play in realising a just and equitable transition to a low-carbon future for South Africa.

The structure of this report is as follows: Section~2 provides background on RL and South Africa’s Just Transition framework. Section~3 presents a detailed motivation for choosing the three key domains of agriculture, energy, and transportation, emphasising their significance within the South African context. Section~4 outlines the methodology used in our systematic literature review. Section~5 summarises the applied RL research we found across these domains. In Section~6 we identify opportunities for extending existing RL work to the South African context and offer a roadmap for future research. Finally, Section~7 addresses the limitations of our work, before we conclude in Section~8.

\section{Background}
To understand the role of RL in South Africa's transition to a low-carbon future, we first provide the necessary background to understand RL and its place within the broader field of AI.
Next, we outline the concept of a JT as defined by the Presidential Climate Commission~\cite{justtransition} and explore the findings of the Presidential Commission on the Fourth Industrial Revolution (4IR)~\cite{gov20204ir}, which emphasises the crucial role that AI will play in achieving South Africa's JT. 

\subsection{Artificial Intelligence and Reinforcement Learning}
AI is a field of computer science that aims to create systems capable of carrying out tasks typically requiring human-like intelligence, such as reasoning, learning, and problem-solving. It encompasses various approaches, including Symbolic AI, which uses explicit rules and symbols for logical reasoning; Expert Systems, which emulate human expertise to make decisions; and Machine Learning (ML), which enables systems to improve over time simply by learning from data.

In recent years, ML has become one of the most prominent branches of AI thanks to significant breakthroughs in Deep Learning (DL) \cite{Goodfellow2016}. Whereas traditional ML techniques either utilise relatively low-dimensional data or require hand-crafted feature extraction from data, DL models automatically discover intricate patterns and representations in the data through hierarchical layers of Artificial Neural Networks (ANNs). Training large DL models on massive amounts of training data has proved to be a powerful recipe for developing programs with surprising levels of intelligence and emergent capabilities \cite{kaplan2020scalinglawsneurallanguage}.

The two most commonly discussed paradigms for training ML models are supervised and unsupervised learning. Supervised learning involves training a model on a labelled dataset, where each input is paired with a corresponding output. Examples include simple regression and classification. Unsupervised learning, on the other hand, deals with unlabelled data. The model is trained to identify underlying patterns or structures in the data without explicit guidance as to the labels underlying these patterns. Examples include clustering and dimensionality reduction.

A third ML paradigm is RL. RL is primarily designed for sequential decision-making problems, whereby an agent must choose actions in an environment, over time, to successfully complete a task. After every action, the agent receives a reward from the environment, which serves as feedback to the agent on how well it is currently doing at the task. As such, RL is evaluative and not instructive, as the agent is not informed what the best action would have been. The agent's goal is to learn a \textit{policy} for selecting actions which result in as much reward as possible over the course of the task. RL algorithms learn through \textit{trial-and-error}, whereby the agent initially chooses actions at random but progressively, throughout training, makes more informed decisions as it learns what the expected reward and consequences of taking actions are.

Before the DL era, RL enjoyed a rich history of research interest \citep{sutton2018reinforcement}, but the scale of problems that could be tackled was limited \citep{kaelbling1996reinforcement}. However, after Mnih et al. \cite{mnih2015dqn} first successfully used Q-learning, an RL algorithm, to train an ANN to play Atari video games from raw pixel values, there was an explosion of \textit{deep} RL (DRL) research and breakthroughs \citep{li2018deepreinforcementlearningoverview}. Most notably, DRL algorithms defeated human world-champions at the famous board game \textit{GO }\citep{silver2017mastering}, and the e-sport video game \textit{DOTA 2} \citep{openai2019dota2largescale}. Since these early successes in DRL on games, the goal has been to improve the generalisability of algorithms and apply them to more complex, real-world problems \citep{degrave2022magnetic}.

Applying RL to real-world problems has proven to be difficult for several reasons. One of the primary challenges is that RL typically requires an agent to interact with its environment many millions of times, which can be impractical in real-world settings. To address this, practitioners often rely on efficient simulators. However, there are limited assurances that policies learned in simulation will effectively transfer to real-world environments~\cite{zhao2020sim}. Additionally, designing a simulator can be a complex task. An alternative approach is to use pre-collected real-world data directly, rather than relying on a simulator. This method, known as offline RL, has garnered significant research attention in recent years due to its potential to make RL more applicable in real-world scenarios \cite{levine2020offlinereinforcementlearningtutorial}.

\subsection{South Africa's Just Transition and Fourth Industrial Revolution}
Governments worldwide are increasingly acknowledging the harmful effects of climate change and taking active measures to combat them. At the 2015 UN Climate Change Conference in Paris, 196 countries signed a legally binding international treaty aimed at addressing climate change \cite{parisagreement}. South Africa, in particular, is confronting several severe and worsening climate impacts, such as droughts, floods, and extreme weather, which intensify existing challenges like poverty, unemployment, and inequality. To help mitigate this looming crisis and fulfil its international obligations, the Presidential Climate Commission \cite{justtransition} was established to provide guidance and support for a just transition to a low-carbon economy. The commission defined a just transition as follows:

\paragraph{\textbf{Just Transition.}} \textit{A just transition aims to achieve a quality life for all South Africans, in the context of increasing the ability to adapt to the adverse impacts of climate, fostering climate resilience, and reaching net-zero greenhouse gas emissions by 2050, in line with the best available science.
A just transition contributes to the goals of decent work for all, social inclusion, and the eradication of poverty.
A just transition puts people at the centre of decision-making, especially those most impacted, the poor, women, people with disabilities, and the youth-empowering and equipping them for new opportunities of the future.
A just transition builds the resilience of the economy and people through affordable, decentralised, diversely owned renewable energy systems; conservation of natural resources; equitable access to water resources; an environment that is not harmful to one’s health and well-being; and sustainable, equitable, inclusive land-use for all, especially for the most vulnerable \cite{justtransition}}. 

\paragraph{}The Presidential Climate Commission's definition of a just transition underscores the government's commitment to significantly reducing greenhouse gas emissions while ensuring that this process is equitable and inclusive. It emphasises a human-centred approach, aiming to avoid harming communities, such as eliminating livelihoods without providing viable alternatives. This approach is particularly crucial in South Africa, where addressing widespread poverty and inequality is as urgent as tackling climate change.

\paragraph{The Fourth Industrial Revolution.} As the global community grapples with rapid changes in climate, it also finds itself in the midst of a swiftly evolving technological era. This transformation is often described as a Fourth Industrial Revolution (4IR), which mirrors past societal shifts like the Agricultural, Industrial, and Information revolutions. The 4IR is marked by the integration of advanced technologies such as AI, robotics, the Internet of Things (IoT), and big data across various sectors. In South Africa, the 4IR is viewed as a crucial opportunity to tackle socio-economic challenges like unemployment and inequality by promoting digital skills development, improving infrastructure, and fostering a knowledge-based economy. The government, in partnership with the private sector and public institutions, is actively developing policies and initiatives to support this transformation, with the goal of ensuring that the benefits of the 4IR are inclusive and broadly distributed across the nation. To advance this objective, a Presidential Commission on 4IR \cite{gov20204ir} was established, providing key recommendations on how the 4IR can be harnessed to achieve South Africa's climate and economic development goals. Among the various emerging technologies, AI was highlighted as having especially significant potential.

\section{Motivation}
For this systematic literature review, we focus on three application areas that are particularly relevant to South Africa's Just Transition. 
The first domain is agriculture, which plays a crucial role in the country by contributing approximately 3\% of GDP and 6\% of employment \cite{agriculture}. 
Agriculture is also vital for food security in South Africa. 
However, the sector is highly vulnerable to climate change due to shifting rainfall patterns, droughts, and floods. 
The second domain is energy. South Africa’s energy mix heavily relies on an ageing fleet of coal-fired power plants, which is incompatible with a green economy and poses reliability challenges that hinder economic growth. 
Lastly, we examine transportation. 
After energy, transportation is the second-largest global source of greenhouse gas emissions \cite{ritchie2025gg}. 
South Africa faces unique challenges in this sector. 
If RL can be applied to mitigate climate risks and improve efficiency in these sectors, it could play a significant role in advancing the Just Transition.

\subsection{Agriculture}
The JT framework highlights climate change's effects on agricultural production, with impacts especially on farm workers and their communities. Not only does climate change put the livelihoods of farming communities at risk, but it also threatens national food security \citep{masipa2017climatechangefoodsecurity}. An Intergovernmental Panel on Climate Change (IPCC) report from 2007 projected that in sub-Saharan Africa, agricultural productivity was at risk of declining from 21\% to 9\% by 2080 \cite{ipcc2007assessment}. More recently, a departmental paper by the International Monetary Fund reported that climate change, amongst other issues, has intensified food insecurity in sub-Saharan Africa with the number of people suffering from high levels of malnutrition, and unable to meet basic food consumption needs increasing by at least 30\% to 123 million in 2022 (or 12\% of the sub-Saharan Africa population)~\cite{imf2022hunger}. 

Furthermore, South Africa is an arid country and is vulnerable to increased occurrences of droughts due to climate change. 
As an example, between 2015 and 2018 Cape Town experienced its worst drought in recorded history, which was so severe that the it nearly became the first modern city to effectively run out of drinking water~\cite{ct-water-crisis-timemag}. 
In the regions surrounding the city, major enterprises are horticulture, and in certain water catchment areas up to 75\% of the water can be allocated to irrigating crops of nearby farms~\citep{johnston2018westerncapefarmers}. 
The drought highlights the complex interactions between water supply for urban and agricultural uses~\cite{Theron2023}. 
Thus, since irrigating crops is such a large consumer of precious water resources, it needs to be carefully controlled and managed.

The Presidential Commission on the 4IR highlighted the role of emerging technologies in bolstering South Africa's resilience to climate change and improving food security. In particular, the commission highlighted the role that precision agriculture could play in enhancing the use of resources and increasing agricultural yield \cite{gov20204ir}. Precision agriculture involves the combined use of autonomous vehicles, drones, remote sensing and AI to optimise the use of water, fertilisers and pesticides and other inputs to maximise agricultural outputs while minimising negative environmental impacts.

\subsection{Energy}
Energy is naturally one of the most salient dimensions of any climate-related transitionary framework, due to the widely accepted role of fossil fuels in climatic change and the urgency to move to greener sources of power. For South Africa, in particular, it exists as a sensitive topic due to the enormous dependence of the country on coal---a fossil-fuel industry responsible for R150 billion of the country's annual income, employing over 100\,000 people~\cite{statssa2021minining}. Regarding the supply of electrical energy specifically, the topic of \emph{loadshedding}---where parts of the grid are switched off to cope with insufficient electricity supply---is currently highly pertinent in the country. In the first half of 2023 alone, a record-breaking 15\,300 GWh of electricity was shed from the grid, with an immense negative economic impact~\cite{SarbEskomNotes}. While in more recent times the reliability of electricity has improved, the national grid remains fragile \cite{dailymaverick-loadshedding}.

The JT framework~\cite{justtransition} rightly recognises the importance of energy in the transition to a brighter future for all South Africans. The framework cites an array of challenges, risks, and opportunities for the country to seize. The Presidential Commission on 4IR~\cite{gov20204ir} corroborates this importance, stating in their report that regression in energy would be ``the single-biggest threat to South Africa---both in respect of human development and economic growth.'' The primary mitigation strategy proposed for South Africa in both of these reports, in line with what is proposed further abroad, centres around the transition to \emph{renewable} energy supplies. With these alternative sources of energy becoming cheaper and more widely accessible, this direction of development is becoming increasingly viable~\cite{akinbami2021state}. There has been early success in this domain, particularly at the metropolitan level, such as in the City of Cape Town, which has been pioneering in their long-term strategy. For example, consider the city's \emph{ENERGY2040} report~\cite{CoCT2040}, which enumerates a host of pathways proposed to overcome a looming energy crisis, many of which are focused around renewable energy sources.

\subsection{Transportation}
After energy generation, transportation is the second-largest source of greenhouse gas emissions globally \cite{ritchie2025gg}. To address this, the global community is increasingly shifting towards more efficient public transport systems. A report by the European Environment Agency found that rail and waterborne transport are significantly more efficient in terms of greenhouse gas emissions than road transport and aviation, both for passenger travel and freight \cite{eea2020trainvsplane}. However, in South Africa, poor maintenance and operational mismanagement of rail infrastructure have led to a decline in rail usage for freight. As a result, the country’s roads have become increasingly burdened with large, inefficient, and often dangerous trucks transporting cargo for export \cite{theeconomist2023rail}. This decline in rail adoption is not only costing the South African economy billions but also exacerbating environmental degradation. To successfully achieve its Just Transition goals, South Africa's public and private sectors are urgently working to rebuild the country’s rail infrastructure and enhance its operational efficiency.

While the transportation of goods is crucial for the JT, the transportation of people is equally essential. The Presidential Climate Commission's latest report on "The State of Climate Action in South Africa" highlights that the public transport system currently faces significant challenges related to accessibility, affordability, and safety. Additionally, both the workers and infrastructure within this sector are highly vulnerable to the impacts of climate change \cite{pcc2024stateofclimateaction}. In South Africa, and across Africa more broadly, extreme weather events driven by climate change pose a serious threat, often causing extensive structural damage to transportation infrastructure and disrupting already fragile and sparse networks \cite{amani2021climaterisktransport}. To address these growing challenges, exacerbated by the climate crisis, cities worldwide are increasingly moving towards becoming "Smart Cities," including some in South Africa~\cite{Paida2016smartcityct,dillip2020bloemfontein}. These interconnected and digitalized urban areas aim to enhance efficiency and resilience through the adoption of emerging technologies like AI.

\section{Research Methodology}
The goal of this systematic literature review is to survey the relevant technical RL literature that is applied to problems related to the three domains of interest; agriculture, energy and transportation.
The search strategy employed was a combination of boolean keyword-based search and citation chaining. 
First, we used Google Scholar to search for papers that contained a combination of keywords we were interested in. 
For example "reinforcement learning" AND ("energy" OR "transportation").
We then checked the reference lists of the papers we found and reviewed the relevant papers they cited.

 
\section{Reinforcement Learning Applications}
Despite the considerable challenges, the application of RL to complex real-world decision-making problems holds immense promise and remains a central focus of research within the AI community. We believe that RL has the potential to play a significant role in South Africa's Just Transition. To support this assertion, we explore three critical domains essential to the Just Transition: agriculture, energy, and transportation. We examine the specific context of these challenges in South Africa and then survey related RL research that could be adapted to the South African context, potentially leading to substantial and tangible impacts.

\subsection{Agriculture} The role of RL in precision agriculture is a nascent area of research which is steadily growing~\cite{GAUTRON2022107182}. Recent works studied the application of RL to learn optimal control for a greenhouse to optimise yields~\cite{Gandhi2022,binas2019reinforcement}. By adjusting greenhouse parameters such as temperature and humidity the growing environment for crops can be optimised and harvest yields maximised~\cite{hemming2020yield}. The challenge with doing RL in a \emph{real-world} environment is that it is very slow to collect training data and we must therefore rely on historical data instead of online interactions. To address this,  researchers have designed simulators for various agriculture decision making tasks including \texttt{gym-DSSAT}~\cite{gautron2022gymdssatcropmodelturned}, \texttt{SWATgym}~\cite{madondo2023swat}, and \texttt{crop-gym}~\cite{overweg2021cropgymreinforcementlearningenvironment}. The recent availability of environment simulators for agriculture makes training and testing RL algorithms far more tractable, and opens up new and exciting avenues of research. Some researchers have studied the application of an RL agent that makes decisions about when to irrigate crop fields based on real-time environmental measurements such as weather patterns. Results have shown significant water-saving potential~\cite{sun2017irrigation,CHEN2021106838,ding2024irrigation,kelly2024irrigation}. Developing similar solutions for the South African context could significantly bolster resilience to future droughts and the availability of irrigation environment simulators such as \texttt{aquacrop-gym}~\cite{kelly2024irrigation} significantly lowers the barrier to entry for researchers. 

\subsection{Energy}
The Presidential Commission on 4IR~\cite{gov20204ir} is notably optimistic about emerging technologies like AI for the application to modern energy solutions. Specifically, there is ample discussion around \emph{smart electrical grids}, with emphasis on novel architectures based on micro-grids and virtual power plants (VPPs). Here, the focus is on solutions that can \emph{dynamically} manage all aspects of the electricity network---the supply, transmission, and distribution of electricity---by finding algorithms which yield optimal control policies. RL has widely been cited as a paradigm highly suitable for this use-case~\cite{ieee_rl_power}, and there have been many promising efforts in this direction. As a demonstrative example, on the generation side, substantial work has gone into the `unit commitment problem' with RL, which aims to coordinate many electrical generators of various capacities, physical dynamics, geographical locations, and costs~\cite{de2022reinforcement}. For transmission and distribution, investigation has been done into responsive voltage control~\cite{ieee_rl_transmission}, and topology reconfiguration or optimisation using RL~\cite{dorfer2022power,kundavcina2022solving}. These are simply instances of energy control using RL, but many directions exist in this vein.

\subsection{Transportation and Logistics}
There has been a significant amount of applied research on ML for optimising freight and logistics operations \cite{Kalliopi2023freight}. Key problem areas include demand forecasting, arrival time prediction, vehicle routing, and traffic flow optimisation, all of which hold considerable potential for the application of RL. For instance, Chen et al.~\cite{chen2023deepfreightintegratingdeepreinforcement} developed a deep RL model called \texttt{DeepFreight}, designed to optimise vehicle dispatching within a fleet to fulfill delivery requests as efficiently as possible. In another study, researchers examined the optimal routing of trucks transferring cargo between port terminals~\cite{Taufik2020itt}, a particularly relevant issue in South Africa, where port inefficiency is a well-documented challenge~\cite{theworldbank2023ports}. Additionally, the lack of timely maintenance of transportation infrastructure presents a significant hurdle in South Africa. Hamida et al.~\cite{Hamida2023infraplanner} explored the use of RL for infrastructure maintenance planning, leveraging hierarchical RL in a simulator they developed called \texttt{InfraPlanner}.

Transportation optimisation represents another promising area for RL research~\cite{Haydari2022transport}. The \texttt{Flatland} challenge~\cite{mohanty2020flatland}, a well-known train routing environment, has garnered considerable research attention. Researchers have also explored using RL to optimise public bus routes~\cite{ahmed2020bus,Sunhyung2023bus}, which could be invaluable in addressing transportation bottlenecks in South Africa’s rapidly growing cities~\cite{Hirsch2023myciti}. Another promising approach to improving urban transportation is through smart traffic light management to reduce congestion. The \texttt{SUMO-RL}~\cite{sumorl} environment, a popular simulator for this purpose, has been widely adopted by the research community~\cite{Hrishit2022trafficlights}.

\begin{table}[t]
\centering
\caption{Open-source RL environments across the three domains of agriculture, energy and transportation.}
\label{tab:simulators}
\vspace{1em}
\scalebox{0.55}{ 
\begin{tabular}{ccll}
\hline
\textbf{Domain}                                                                        & \textbf{Environment Name} & \multicolumn{1}{c}{\textbf{Description}}                                                                                                                                                                                                                              & \multicolumn{1}{c}{\textbf{Link}}                         \\ \hline
\multirow{4}{*}{Agriculture}                                                           & aquacrop-gym              & \begin{tabular}[c]{@{}l@{}}An RL gym environment to train and compare \\ irrigation scheduling strategies.\end{tabular}                                                                                                                                               & \url{https://github.com/aquacropos/aquacrop-gym}                \\ \cline{2-4} 
                                                                                       & CropGym                   & \begin{tabular}[c]{@{}l@{}}An RL gym environment where a RL agent \\ can learn fertilization management policies.\end{tabular}                                                                                                                                        & \url{https://github.com/BigDataWUR/crop-gym}                    \\ \cline{2-4} 
                                                                                       & gym-DSSAT                 & \begin{tabular}[c]{@{}l@{}}A modification of the Decision Support \\ System for Agrotechnology Transfer (DSSAT) \\ Fortran software into an easy-to-manipulate Python \\ RL gym environment for researchers.\end{tabular}                                             & \url{https://gitlab.inria.fr/rgautron/gym\_dssat\_pdi}          \\ \cline{2-4} 
                                                                                       & SWATgym                   & \begin{tabular}[c]{@{}l@{}}A RL environment based on the Soil and \\ Water Assessment Tool (SWAT). SWAT is a physics \\ based river basin model that has been widely used to \\ evaluate the effects of crop management decisions on \\ water resources.\end{tabular} & \url{https://github.com/IBM/SWATgym}                            \\ \hline
\multirow{5}{*}{Energy}                                                                & CityLearn                 & \begin{tabular}[c]{@{}l@{}}A RL gym environment for the \\ implementation RL for building energy coordination \\ and demand response in cities.\end{tabular}                                                                                                          & \url{https://github.com/intelligent-environments-lab/CityLearn} \\ \cline{2-4} 
                                                                                       & CommonPower               & \begin{tabular}[c]{@{}l@{}}A flexible framework to model power systems and \\ benchmark safe controllers.\end{tabular}                                                                                                                                                & \url{https://github.com/TUMcps/commonpower}                     \\ \cline{2-4} 
                                                                                       & SustainGym                & \begin{tabular}[c]{@{}l@{}}A suite of five environments designed to test the \\ performance of RL algorithms on realistic sustainable \\ energy system tasks, ranging from electric vehicle \\ charging to carbon-aware data-centre job scheduling.\end{tabular}      & \url{https://github.com/chrisyeh96/sustaingym}                  \\ \cline{2-4} 
                                                                                       & RL4UC                     & RL environment for unit commitment.                                                                                                                                                                                                                                   & \url{https://github.com/pwdemars/rl4uc}                         \\ \cline{2-4} 
                                                                                       & Grid2Op                   & \begin{tabular}[c]{@{}l@{}}RL environment to model sequential decision \\ making on power grids.\end{tabular}                                                                                                                                                         & \url{https://github.com/rte-france/Grid2Op}                     \\ \hline
\multirow{4}{*}{\begin{tabular}[c]{@{}c@{}}Transportation\\ \& Logistics\end{tabular}} & Flatland                  & \begin{tabular}[c]{@{}l@{}}An RL environment aiming to test RL \\ solutions for scheduling trains on large, procedurally \\ generated networks.\end{tabular}                                                                                                          & \url{https://gitlab.aicrowd.com/flatland/flatland}              \\ \cline{2-4} 
                                                                                       & DeepFreight               & A freight delivery scheduling environment.                                                                                                                                                                                                                            & \url{https://github.com/LucasCJYSDL/DeepFreight}                \\ \cline{2-4} 
                                                                                       & InfraPlanner              & \begin{tabular}[c]{@{}l@{}}Benchmark RL environment for infrastructure \\ maintenance planning\end{tabular}                                                                                                                                                           & \url{https://github.com/CivML-PolyMtl/InfrastructuresPlanner}   \\ \cline{2-4} 
                                                                                       & SUMO-RL                   & \begin{tabular}[c]{@{}l@{}}RL environments for Traffic Signal Control \\ with SUMO.\end{tabular}                                                                                                                                                                      & \url{https://github.com/LucasAlegre/sumo-rl}                    \\ \hline
\end{tabular}
}
\end{table}

\section{Roadmap for Future Researchers}
With our three focus areas---agriculture, energy, and transportation---explained and synergised with contemporary RL literature, what is the road ahead? In this paper, we aim to offer two resources to the broader community. Firstly, we present a collection of available resources for researchers---including the previously cited papers as examples, along with many open-source codebases. The aim is for these to orient new-joiners and help them get started in the field. Secondly, we taxonomise three \emph{themes} of work, with associated open questions, where we feel there will be maximal positive impact. We centre our work on the South African context specifically.

\subsection{Building on the resources available}
Previously, we outlined a wide range of works that attempt to apply RL to real-world tasks. Researchers interested in applying RL in the South African context should certainly start by building on these foundations, rather than necessarily starting from scratch. In Table~\ref{tab:simulators} we summarise the wide range of environment simulators that we found relevant to the problems discussed. 

\subsection{Open problems and future work}
Now, we outline three distinct themes of work going forward, along with associated open problems.

\paragraph{\textbf{Theme 1: Designing Simulators for South African Problem Settings.}} Building on the existing work that has contributed RL environment simulators for various problem settings, there is a significant opportunity for researchers to develop RL environment simulators tailored to address South Africa’s unique challenges. By creating simulators that reflect the specific conditions and needs of the country, researchers can make substantial contributions to the effectiveness and applicability of RL in the local context. Here are three examples of where such simulators could be invaluable:

\begin{itemize}
    \item \textbf{Irrigation and Crop Simulators:} Developing simulators that incorporate South Africa’s diverse crop types and regional rainfall patterns would be highly beneficial. These simulators could model the effects of the country's variable climate, which includes distinct rainfall seasons—summer rainfall in the interior and winter rainfall in the Cape—providing a more accurate tool for optimising agricultural practices.

    \item \textbf{Power Network Management Simulators:} A simulator that accounts for the unique challenges of South Africa’s power grid could greatly enhance RL research in this area. This includes the frequent power outages (load shedding) that are far more common than in more developed economies, the long distances over which power must be transmitted, and the country’s varied climate, which includes areas with very high solar irradiance and/or wind. Such a simulator could aid in developing strategies for more resilient and efficient energy distribution.

    \item \textbf{Transportation Network Simulators:} Creating simulators that reflect the unique challenges of South African cities could address critical transportation issues. The legacy of apartheid has left a lasting impact on city layouts, resulting in long, sparse commuting corridors that are easily disrupted. These simulators could model the inefficiencies and inequities of the current transport system, including the challenges posed by the minibus taxi industry, which fills some gaps but also faces issues of safety and affordability. Additionally, simulators could explore the impact of South Africa’s extensive but ageing rail network, which, despite being one of the largest in sub-Saharan Africa, suffers from neglect and requires significant upkeep. Developing such simulators could help in creating more equitable and efficient transportation solutions that are vital for the country’s Just Transition.
\end{itemize}

\paragraph{\textbf{Theme 2: Collating South African data for Offline RL.}} Offline Reinforcement Learning (RL) is an alternative approach to traditional RL that does not rely on a simulator \cite{levine2020offlinereinforcementlearningtutorial}. Instead, it trains models directly on pre-collected experience, making it a promising direction for applying RL in real-world scenarios where building a simulator would be impractical or prohibitively expensive. However, for offline RL to be effective, researchers and practitioners need access to extensive and relevant training data. Currently, no offline RL datasets are available for South African-specific problem settings, which presents a significant opportunity. Collecting and making the following types of data accessible to the research community could greatly enhance the applicability of offline RL in South Africa:

\begin{itemize}
    \item \textbf{Agricultural Data:} Historical data on crop yields and weather patterns would be invaluable for developing offline RL systems that can support more informed and effective agricultural decision-making, particularly in the context of South Africa’s diverse climates and farming practices.

    \item \textbf{Electricity Usage Data:} Access to detailed electricity consumption data could enable the development of offline RL models for better load balancing and automatic demand response, which are critical for improving the reliability and sustainability of South Africa's power grid.

    \item \textbf{Transportation Usage Data:} Data from public transport systems, such as the MyCiTi bus service, could be used to design offline RL systems for optimising bus schedules, improving efficiency, and better meeting passenger needs.
\end{itemize}

By focusing on these areas, researchers can harness the power of offline RL to address some of South Africa's most pressing challenges, driving progress in sectors vital to the country's Just Transition.

\paragraph{\textbf{Theme 3: Fundamental research on simulator-to-real transfer.}} A major challenge in applying Reinforcement Learning (RL) to real-world scenarios is the uncertainty of whether policies learned in simulated environments will perform effectively, in reality, \cite{Salvato2021sim2real,wenshuaisim2real}. This "simulator-to-real transfer" problem limits the practical adoption of RL, as there are no guarantees that strategies optimised in simulations will succeed when deployed outside them. To address this issue, fundamental research is needed to bridge the gap between simulation and real-world applications. This could involve developing more accurate simulators, creating algorithms that are more adaptable to real-world variations, or exploring techniques like domain adaptation and transfer learning. Progress in this area is essential for making RL widely applicable in industries such as autonomous systems and smart energy management. Without it, the full potential of RL in real-world settings will remain unrealised, in South Africa and beyond.

\section{Limitations}
Crucially, this work should not be taken as a panacea to the problems facing South Africa---a simple one-shot fix. These efforts are just one component of a broader suite of solutions to achieving the goals of a future that is truly just and equitable. Technology plays an important part of in this transition, indeed, but it cannot exist without other innovations and collaborations across a range of disciplines and stakeholders. Moreover, even within technological development, there are a range of factors to consider. For example, we have listed some (but not all) of the other dimensions worth noting in these discussions, below.
\begin{itemize}
    \item \textbf{Digital Divide.} The benefits of AI and 4IR are not equally distributed, and there is a risk of widening the digital divide between those who have access to technology and those who do not. Efforts are needed to ensure that all South Africans can participate in and benefit from the 4IR.
    \item \textbf{Ethical Concerns.} The deployment of AI raises ethical issues, including data privacy, bias in AI algorithms, and the potential for job displacement. How should we handle theses concerns and the affected people in a just way? Addressing them is crucial for the responsible development and use of AI. 
    \item \textbf{Infrastructure and Investment.} For South Africa to fully leverage AI, significant investment in digital infrastructure, such as high-speed internet and data centres, is required. Additionally, fostering an innovation-friendly environment through supportive policies and investment in research and development is essential.
\end{itemize}

Ultimately, we feel that developments towards a better future should always remain human-centred and multi-faceted. There are many barriers still to be overcome, but if we can pursue these ambitious goals with passion and purpose, a brighter future is certainly possible.

\section{Conclusion}
In this paper, we have demonstrated that there is a significant body of applied RL work in the three critical domains of agriculture, energy, and transportation. 
These domains are essential for advancing South Africa's Just Transition towards a more sustainable and inclusive future. 
However, despite the progress in applying RL to these areas, there are currently no significant RL applications specifically tailored to the South African context. 
This gap presents an exciting opportunity for future researchers to adapt and implement RL solutions that address the unique challenges faced by South Africa. 
To this end, we have provided a roadmap outlining key areas where RL can be applied and emphasised the importance of developing context-specific tools and strategies. 
By building on existing research and addressing local needs, future researchers can make substantial contributions to the Just Transition, ensuring that it is both technologically advanced and socially equitable.

\begin{credits}

\subsubsection{\discintname}
The authors have no competing interests to declare.
\end{credits}
%
%
%
\bibliographystyle{plainnat}
\bibliography{main}

\begin{thebibliography}{64}
\providecommand{\natexlab}[1]{#1}
\providecommand{\url}[1]{\texttt{#1}}
\expandafter\ifx\csname urlstyle\endcsname\relax
  \providecommand{\doi}[1]{doi: #1}\else
  \providecommand{\doi}{doi: \begingroup \urlstyle{rm}\Url}\fi

\bibitem[Akinbami et~al.(2021)Akinbami, Oke, and Bodunrin]{akinbami2021state}
Olusola~M Akinbami, Samuel~R Oke, and Michael~O Bodunrin.
\newblock The state of renewable energy development in south africa: An overview.
\newblock \emph{Alexandria Engineering Journal}, 2021.

\bibitem[Alegre(2019)]{sumorl}
Lucas~N. Alegre.
\newblock Sumo-rl.
\newblock \emph{GitHub repository}, 2019.

\bibitem[Baker(2020)]{ct-water-crisis-timemag}
Aryn Baker.
\newblock I knew we were in trouble. what it’s like to live through cape town’s massive water crisis.
\newblock \emph{TIME}, 2020.
\newblock URL \url{https://time.com/cape-town-south-africa-water-crisis/}.

\bibitem[Baptista et~al.(2022)]{imf2022hunger}
Diogo Miguel~Salgado Baptista et~al.
\newblock Climate change and chronic food insecurity in sub-saharan africa.
\newblock \emph{IMF Departmental Papers}, 2022.

\bibitem[Binas et~al.(2019)Binas, Luginbuehl, and Bengio]{binas2019reinforcement}
Jonathan Binas, Leonie Luginbuehl, and Yoshua Bengio.
\newblock Reinforcement learning for sustainable agriculture.
\newblock In \emph{ICML 2019 Workshop on Climate Change: How Can AI Help?}, 2019.
\newblock URL \url{https://www.climatechange.ai/papers/icml2019/32}.

\bibitem[Chaudhuri et~al.(2022)Chaudhuri, Masti, Veerendranath, and Natarajan]{Hrishit2022trafficlights}
Hrishit Chaudhuri, Vibha Masti, Vishruth Veerendranath, and S.~Natarajan.
\newblock A comparative study of algorithms for intelligent traffic signal control.
\newblock \emph{Machine Learning and Autonomous Systems}, 2022.

\bibitem[Chen et~al.(2023)Chen, Umrawal, Lan, and Aggarwal]{chen2023deepfreightintegratingdeepreinforcement}
Jiayu Chen, Abhishek~K. Umrawal, Tian Lan, and Vaneet Aggarwal.
\newblock Deepfreight: Integrating deep reinforcement learning and mixed integer programming for multi-transfer truck freight delivery, 2023.

\bibitem[Chen et~al.(2021)Chen, Cui, Wang, Xie, Liu, Luo, Zheng, and Luo]{CHEN2021106838}
Mengting Chen, Yuanlai Cui, Xiaonan Wang, Hengwang Xie, Fangping Liu, Tongyuan Luo, Shizong Zheng, and Yufeng Luo.
\newblock A reinforcement learning approach to irrigation decision-making for rice using weather forecasts.
\newblock \emph{Agricultural Water Management}, 2021.

\bibitem[CoCT(2015)]{CoCT2040}
CoCT.
\newblock Cape town energy2040.
\newblock Technical report, City of Cape Town (CoCT), 2015.

\bibitem[DALRRD(2022)]{agriculture}
DALRRD.
\newblock Economic review of the south african agriculture.
\newblock Technical report, Department of Agriculture, land reform and rural development (DALRRD), 2022.

\bibitem[Darwish et~al.(2020)Darwish, Khalil, and Badawi]{ahmed2020bus}
Ahmed Darwish, Momen Khalil, and Karim Badawi.
\newblock Optimising public bus transit networks using deep reinforcement learning.
\newblock \emph{IEEE 23rd International Conference on Intelligent Transportation Systems (ITSC)}, 2020.

\bibitem[Das(2020)]{dillip2020bloemfontein}
Dillip~Kumar Das.
\newblock Perspectives of smart cities in south africa through applied systems analysis approach: A case of bloemfontein.
\newblock \emph{Construction Economics and Building}, 2020.

\bibitem[De~Mars(2022)]{de2022reinforcement}
Patrick De~Mars.
\newblock Reinforcement learning and tree search methods for the unit commitment problem.
\newblock \emph{Arxiv Preprint}, 2022.

\bibitem[Degrave et~al.(2022)Degrave, Felici, Buchli, Neunert, Tracey, Carpanese, Ewalds, Hafner, Abdolmaleki, de~Las~Casas, et~al.]{degrave2022magnetic}
Jonas Degrave, Federico Felici, Jonas Buchli, Michael Neunert, Brendan Tracey, Francesco Carpanese, Timo Ewalds, Roland Hafner, Abbas Abdolmaleki, Diego de~Las~Casas, et~al.
\newblock Magnetic control of tokamak plasmas through deep reinforcement learning.
\newblock \emph{Nature}, 2022.

\bibitem[Ding and Du(2024)]{ding2024irrigation}
Xianzhong Ding and Wan Du.
\newblock Optimizing irrigation efficiency using deep reinforcement learning in the field.
\newblock \emph{Association for Computing Machinery}, 2024.

\bibitem[Dorfer et~al.(2022)Dorfer, Fuxj{\"a}ger, Kozak, Blies, and Wasserer]{dorfer2022power}
Matthias Dorfer, Anton~R Fuxj{\"a}ger, Kristian Kozak, Patrick~M Blies, and Marcel Wasserer.
\newblock Power grid congestion management via topology optimization with alphazero.
\newblock \emph{Arxiv Pre Print}, 2022.

\bibitem[Economist(2023)]{theeconomist2023rail}
The Economist.
\newblock South africa’s disintegrating freight railway is crippling firms.
\newblock \emph{The Economist}, 2023.

\bibitem[EEA(2020)]{eea2020trainvsplane}
EEA.
\newblock Transport and environment report 2020, train or plane.
\newblock Technical report, European Environment Agency (EEA), 2020.

\bibitem[Gandhi(2022)]{Gandhi2022}
Ratnik Gandhi.
\newblock \emph{Deep Reinforcement Learning for Agriculture: Principles and Use Cases}.
\newblock Springer, 2022.

\bibitem[Gautron et~al.(2022{\natexlab{a}})Gautron, Maillard, Preux, Corbeels, and Sabbadin]{GAUTRON2022107182}
Romain Gautron, Odalric-Ambrym Maillard, Philippe Preux, Marc Corbeels, and Régis Sabbadin.
\newblock Reinforcement learning for crop management support: Review, prospects and challenges.
\newblock \emph{Computers and Electronics in Agriculture}, 2022{\natexlab{a}}.

\bibitem[Gautron et~al.(2022{\natexlab{b}})Gautron, Padrón, Preux, Bigot, Maillard, and Emukpere]{gautron2022gymdssatcropmodelturned}
Romain Gautron, Emilio~J. Padrón, Philippe Preux, Julien Bigot, Odalric-Ambrym Maillard, and David Emukpere.
\newblock gym-dssat: a crop model turned into a reinforcement learning environment, 2022{\natexlab{b}}.

\bibitem[Goodfellow et~al.(2016)Goodfellow, Bengio, and Courville]{Goodfellow2016}
Ian Goodfellow, Yoshua Bengio, and Aaron Courville.
\newblock \emph{Deep Learning}.
\newblock MIT Press, 2016.
\newblock \url{http://www.deeplearningbook.org}.

\bibitem[Hamida and Goulet(2023)]{Hamida2023infraplanner}
Zachary Hamida and James-A. Goulet.
\newblock Hierarchical reinforcement learning for transportation infrastructure maintenance planning.
\newblock \emph{Reliability Engineering \& System Safety}, 2023.

\bibitem[Haydari and Yılmaz(2022)]{Haydari2022transport}
Ammar Haydari and Yasin Yılmaz.
\newblock Deep reinforcement learning for intelligent transportation systems: A survey.
\newblock \emph{IEEE Transactions on Intelligent Transportation Systems}, 2022.

\bibitem[Hemming et~al.(2020)Hemming, Zwart, Elings, Petropoulou, and Righini]{hemming2020yield}
Silke Hemming, Feije~de Zwart, Anne Elings, Anna Petropoulou, and Isabella Righini.
\newblock Cherry tomato production in intelligent greenhouses—sensors and ai for control of climate, irrigation, crop yield, and quality.
\newblock \emph{Sensors}, 2020.

\bibitem[Hirsch(2023)]{Hirsch2023myciti}
Matthew Hirsch.
\newblock Myciti the focus of cape town’s new public transport business plan.
\newblock \emph{GroundUp}, 2023.

\bibitem[IPCC(2007)]{ipcc2007assessment}
IPCC.
\newblock Intergovernmental panel on climate change fourth assessment report.
\newblock \emph{IPCC}, 2007.

\bibitem[Johnston(2018)]{johnston2018westerncapefarmers}
Peter Johnston.
\newblock How western cape farmers are being hit by the drought.
\newblock \emph{The Conversation}, 2018.

\bibitem[Kaelbling et~al.(1996)Kaelbling, Littman, and Moore]{kaelbling1996reinforcement}
Leslie~Pack Kaelbling, Michael~L Littman, and Andrew~W Moore.
\newblock Reinforcement learning: A survey.
\newblock \emph{Journal of artificial intelligence research}, 1996.

\bibitem[Kaplan et~al.(2020)Kaplan, McCandlish, Henighan, Brown, Chess, Child, Gray, Radford, Wu, and Amodei]{kaplan2020scalinglawsneurallanguage}
Jared Kaplan, Sam McCandlish, Tom Henighan, Tom~B. Brown, Benjamin Chess, Rewon Child, Scott Gray, Alec Radford, Jeffrey Wu, and Dario Amodei.
\newblock Scaling laws for neural language models.
\newblock \emph{ArXiv Preprint}, 2020.

\bibitem[Kelly et~al.(2024)Kelly, Foster, and Schultz]{kelly2024irrigation}
T.D. Kelly, T.~Foster, and David~M. Schultz.
\newblock Assessing the value of deep reinforcement learning for irrigation scheduling.
\newblock \emph{Smart Agricultural Technology}, 2024.

\bibitem[Kunda{\v{c}}ina et~al.(2022)Kunda{\v{c}}ina, Vidovi{\'c}, and Petkovi{\'c}]{kundavcina2022solving}
Ognjen~B Kunda{\v{c}}ina, Predrag~M Vidovi{\'c}, and Milan~R Petkovi{\'c}.
\newblock Solving dynamic distribution network reconfiguration using deep reinforcement learning.
\newblock \emph{Electrical Engineering}, 2022.

\bibitem[Levine et~al.(2020)Levine, Kumar, Tucker, and Fu]{levine2020offlinereinforcementlearningtutorial}
Sergey Levine, Aviral Kumar, George Tucker, and Justin Fu.
\newblock Offline reinforcement learning: Tutorial, review, and perspectives on open problems.
\newblock \emph{Arxiv Preprint}, 2020.

\bibitem[Li(2018)]{li2018deepreinforcementlearningoverview}
Yuxi Li.
\newblock Deep reinforcement learning: An overview.
\newblock \emph{Arxiv Preprint}, 2018.

\bibitem[Madondo et~al.(2023)Madondo, Azmat, Dipietro, Horesh, Jacobs, Bawa, Srinivasan, and O'Donncha]{madondo2023swat}
Malvern Madondo, Muneeza Azmat, Kelsey Dipietro, Raya Horesh, Michael Jacobs, Arun Bawa, Raghavan Srinivasan, and Fearghal O'Donncha.
\newblock A swat-based reinforcement learning framework for crop management.
\newblock \emph{Arxiv Preprint}, 2023.

\bibitem[Masipa(2017)]{masipa2017climatechangefoodsecurity}
Tshepo Masipa.
\newblock The impact of climate change on food security in south africa: Current realities and challenges ahead.
\newblock \emph{Jamba}, 2017.

\bibitem[Mhangara et~al.(2016)Mhangara, Mudau, Mboup, and Mwaniki]{Paida2016smartcityct}
Paida Mhangara, Naledzani Mudau, Gora Mboup, and Dennis Mwaniki.
\newblock Transforming the city of cape town from an apartheid city to an inclusive smart city.
\newblock \emph{Smart Economy in Smart Cities}, 2016.

\bibitem[Mnih et~al.(2015)Mnih, Kavukcuoglu, Silver, Rusu, Veness, Bellemare, Graves, Riedmiller, Fidjeland, Ostrovski, Petersen, Beattie, Sadik, Antonoglou, King, Kumaran, Wierstra, Legg, and Hassabis]{mnih2015dqn}
Volodymyr Mnih, Koray Kavukcuoglu, David Silver, Andrei~A. Rusu, Joel Veness, Marc~G. Bellemare, Alex Graves, Martin Riedmiller, Andreas~K. Fidjeland, Georg Ostrovski, Stig Petersen, Charles Beattie, Amir Sadik, Ioannis Antonoglou, Helen King, Dharshan Kumaran, Daan Wierstra, Shane Legg, and Demis Hassabis.
\newblock Human-level control through deep reinforcement learning.
\newblock \emph{Nature}, 2015.

\bibitem[Mohanty et~al.(2020)Mohanty, Nygren, Laurent, Schneider, Scheller, Bhattacharya, Watson, Egli, Eichenberger, Baumberger, et~al.]{mohanty2020flatland}
Sharada Mohanty, Erik Nygren, Florian Laurent, Manuel Schneider, Christian Scheller, Nilabha Bhattacharya, Jeremy Watson, Adrian Egli, Christian Eichenberger, Christian Baumberger, et~al.
\newblock Flatland-rl: Multi-agent reinforcement learning on trains.
\newblock \emph{ArXiv Preprint}, 2020.

\bibitem[OpenAI et~al.(2019)OpenAI, :, Berner, Brockman, Chan, Cheung, Dębiak, Dennison, Farhi, Fischer, Hashme, Hesse, Józefowicz, Gray, Olsson, Pachocki, Petrov, d.~O.~Pinto, Raiman, Salimans, Schlatter, Schneider, Sidor, Sutskever, Tang, Wolski, and Zhang]{openai2019dota2largescale}
OpenAI, :, Christopher Berner, Greg Brockman, Brooke Chan, Vicki Cheung, Przemysław Dębiak, Christy Dennison, David Farhi, Quirin Fischer, Shariq Hashme, Chris Hesse, Rafal Józefowicz, Scott Gray, Catherine Olsson, Jakub Pachocki, Michael Petrov, Henrique~P. d.~O.~Pinto, Jonathan Raiman, Tim Salimans, Jeremy Schlatter, Jonas Schneider, Szymon Sidor, Ilya Sutskever, Jie Tang, Filip Wolski, and Susan Zhang.
\newblock Dota 2 with large scale deep reinforcement learning.
\newblock \emph{ArXiv Preprint}, 2019.

\bibitem[Overweg et~al.(2021)Overweg, Berghuijs, and Athanasiadis]{overweg2021cropgymreinforcementlearningenvironment}
Hiske Overweg, Herman N.~C. Berghuijs, and Ioannis~N. Athanasiadis.
\newblock Cropgym: a reinforcement learning environment for crop management, 2021.

\bibitem[PCC(2022)]{justtransition}
PCC.
\newblock A framework for a just transition in south africa.
\newblock Technical report, Presidential Climate Commission (PCC), 2022.

\bibitem[PCC(2024)]{pcc2024stateofclimateaction}
PCC.
\newblock The state of climate action in south africa.
\newblock \emph{Presidential Climate Comission}, 2024.

\bibitem[Ramesh et~al.(2021)Ramesh, Pavlov, Goh, Gray, Voss, Radford, Chen, and Sutskever]{ramesh2021zeroshottexttoimagegeneration}
Aditya Ramesh, Mikhail Pavlov, Gabriel Goh, Scott Gray, Chelsea Voss, Alec Radford, Mark Chen, and Ilya Sutskever.
\newblock Zero-shot text-to-image generation.
\newblock \emph{ArXiv Preprint}, 2021.

\bibitem[Revolution(2020)]{gov20204ir}
Presidential Commission On The Fourth~Industrial Revolution.
\newblock Summary report and recommendations.
\newblock \emph{South Africa}, 2020.

\bibitem[Ritchie et~al.(2024)Ritchie, Rosado, and Roser]{ritchie2025gg}
Hannah Ritchie, Pablo Rosado, and Max Roser.
\newblock Breakdown of carbon dioxide, methane and nitrous oxide emissions by sector.
\newblock \emph{Our World in Data}, 2024.

\bibitem[Rweyendela(2021)]{amani2021climaterisktransport}
Amani~George Rweyendela.
\newblock Climate change is a threat to africa’s transport systems: what must be done.
\newblock \emph{The Conversation}, 2021.

\bibitem[Salvato et~al.(2021)Salvato, Fenu, Medvet, and Pellegrino]{Salvato2021sim2real}
Erica Salvato, Gianfranco Fenu, Eric Medvet, and Felice~Andrea Pellegrino.
\newblock Crossing the reality gap: A survey on sim-to-real transferability of robot controllers in reinforcement learning.
\newblock \emph{IEEE Access}, 2021.

\bibitem[SARB(2024)]{SarbEskomNotes}
SARB.
\newblock South african reserve bank occasional bulletin of economic notes.
\newblock Technical report, SARB, April 2024.

\bibitem[Sibanyoni and Comrie(2024)]{dailymaverick-loadshedding}
Buyeleni Sibanyoni and Susan Comrie.
\newblock Here’s the data showing how eskom is able to keep sa’s lights on.
\newblock \emph{amaBhungane}, 2024.
\newblock URL \url{https://www.dailymaverick.co.za/article/2024-10-01-heres-the-data-showing-how-eskom-is-able-to-keep-sas-lights-on/}.

\bibitem[Silver et~al.(2017)Silver, Schrittwieser, Simonyan, Antonoglou, Huang, Guez, Hubert, Baker, Lai, Bolton, et~al.]{silver2017mastering}
David Silver, Julian Schrittwieser, Karen Simonyan, Ioannis Antonoglou, Aja Huang, Arthur Guez, Thomas Hubert, Lucas Baker, Matthew Lai, Adrian Bolton, et~al.
\newblock Mastering the game of go without human knowledge.
\newblock \emph{Nature}, 2017.

\bibitem[StatsSA(2019)]{statssa2021minining}
StatsSA.
\newblock Mining industry, 2019.
\newblock Technical report, StatsSA, 2019.
\newblock URL \url{https://www.statssa.gov.za/publications/Report-20-01-02/Report-20-01-022019.pdf}.

\bibitem[Sun et~al.(2017)Sun, Yang, Hu, Porter, Marek, and Hillyer]{sun2017irrigation}
Lijia Sun, Yanxiang Yang, Jiang Hu, Dana Porter, Thomas Marek, and Charles Hillyer.
\newblock Reinforcement learning control for water-efficient agricultural irrigation.
\newblock In \emph{IEEE International Conference on Ubiquitous Computing and Communications}, 2017.

\bibitem[Sutton and Barto(2018)]{sutton2018reinforcement}
Richard~S Sutton and Andrew~G Barto.
\newblock \emph{Reinforcement learning: An introduction}.
\newblock Online, 2018.

\bibitem[Taufik Nur~Adi and Bae(2020)]{Taufik2020itt}
Yelita Anggiane~Iskandar Taufik Nur~Adi and Hyerim Bae.
\newblock Interterminal truck routing optimization using deep reinforcement learning.
\newblock \emph{Sensors}, 2020.

\bibitem[Thayer and Overbye(2020)]{ieee_rl_transmission}
Brandon~L. Thayer and Thomas~J. Overbye.
\newblock Deep reinforcement learning for electric transmission voltage control.
\newblock In \emph{2020 IEEE Electric Power and Energy Conference (EPEC)}, 2020.

\bibitem[Theron et~al.(2023)Theron, Midgley, Hochrainer-Stigler, Archer, Tramberand, and Walker]{Theron2023}
Simone~Norah Theron, Stephanie Midgley, Stefan Hochrainer-Stigler, Emma Archer, Sylvia Tramberand, and Sue Walker.
\newblock Agricultural resilience and adaptive capacity during severe drought in the western cape, south africa.
\newblock \emph{Regional Environmental Change}, 2023.

\bibitem[Tsolaki et~al.(2023)Tsolaki, Vafeiadis, Nizamis, Ioannidis, and Tzovaras]{Kalliopi2023freight}
Kalliopi Tsolaki, Thanasis Vafeiadis, Alexandros Nizamis, Dimosthenis Ioannidis, and Dimitrios Tzovaras.
\newblock Utilizing machine learning on freight transportation and logistics applications: A review.
\newblock \emph{ICT Express}, 2023.

\bibitem[UN(2015)]{parisagreement}
UN.
\newblock The paris agreement.
\newblock Technical report, United Nations (UN) Framework Convention on Climate Change, 2015.

\bibitem[WB(2023)]{theworldbank2023ports}
WB.
\newblock The container port performance index.
\newblock Technical report, The World Bank (WB), 2023.

\bibitem[Yoo and Lee(2023)]{Sunhyung2023bus}
Sunhyung Yoo and Jinwoo~(Brian) Lee.
\newblock Revising bus routes to improve access for the transport disadvantaged: A reinforcement learning approach.
\newblock \emph{Journal of Public Transportation}, 2023.

\bibitem[Zhang et~al.(2018)Zhang, Han, and Deng]{ieee_rl_power}
Dongxia Zhang, Xiaoqing Han, and Chunyu Deng.
\newblock Review on the research and practice of deep learning and reinforcement learning in smart grids.
\newblock \emph{CSEE Journal of Power and Energy Systems}, 2018.

\bibitem[Zhao et~al.(2020{\natexlab{a}})Zhao, Queralta, and Westerlund]{wenshuaisim2real}
Wenshuai Zhao, Jorge~Pe{\~{n}}a Queralta, and Tomi Westerlund.
\newblock Sim-to-real transfer in deep reinforcement learning for robotics: a survey.
\newblock \emph{CoRR}, 2020{\natexlab{a}}.

\bibitem[Zhao et~al.(2020{\natexlab{b}})Zhao, Queralta, and Westerlund]{zhao2020sim}
Wenshuai Zhao, Jorge~Pe{\~n}a Queralta, and Tomi Westerlund.
\newblock Sim-to-real transfer in deep reinforcement learning for robotics: a survey.
\newblock In \emph{2020 IEEE symposium series on computational intelligence (SSCI)}. IEEE, 2020{\natexlab{b}}.

\end{thebibliography}
\end{document}